\documentclass[useAMS,usenatbib]{mn2e}
\usepackage{txfonts}
\usepackage{graphicx}
\usepackage{multirow}
\usepackage{dcolumn}
\usepackage{array}
\usepackage{xcolor}
\usepackage{float}

\newcommand{\PreserveBackslash}[1]{\let\temp=\\#1\let\\=\temp}
\newcolumntype{C}[1]{>{\PreserveBackslash\centering}p{#1}}
\newcolumntype{R}[1]{>{\PreserveBackslash\raggedleft}p{#1}}
\newcolumntype{L}[1]{>{\PreserveBackslash\raggedright}p{#1}}

\title[]{Thermophysical characteristics of the large main-belt asteroid (349) Dembowska}

\author[LiangLiang Yu et al.]{LiangLiang Yu$^{1,2}$\thanks{yullmoon@live.com}, Bin Yang$^{3,4}$,
Jianghui Ji$^{2}$\thanks{jijh@pmo.ac.cn}, Wing-Huen Ip$^{1,5}$\\
$^{1}$Space Science Institute, Macau University of Science and Technology, Taipa, Macau; \\
$^{2}$CAS Key Laboratory of Planetary Sciences, Purple Mountain Observatory, Chinese Academy of Sciences, Nanjing 210008, China; \\
$^{3}$European Southern Observatory, Alonso de C\'{o}rdova 3107, Vitacura, Casilla 19001, Santiago, Chile; \\
$^{4}$Yunnan Observatories, Chinese Academy of Sciences, Kunming 650011, China; \\
$^{5}$Institute of Astronomy, National Central University, Jhongli, Taoyuan City 32001, Taiwan}

\begin{document}
\date{Received 2014 December 14; in original form 2014 December 30}

\pagerange{\pageref{firstpage}--\pageref{lastpage}} \pubyear{2002}

\maketitle

\label{firstpage}

\begin{abstract}
(349) Dembowska, a large, bright main-belt asteroid, has a fast rotation and
oblique spin axis. It may have experienced partial melting and differentiation.
We constrain Dembowska's thermophysical properties, e.g., thermal inertia, roughness
fraction, geometric albedo and effective diameter within 3$\sigma$ uncertainty of
$\Gamma=20^{+12}_{-7}\rm~Jm^{-2}s^{-0.5}K^{-1}$, $f_{\rm r}=0.25^{+0.60}_{-0.25}$,
$p_{\rm v}=0.309^{+0.026}_{-0.038}$, and $D_{\rm eff}=155.8^{+7.5}_{-6.2}\rm~km$,
by utilizing the Advanced Thermophysical Model (ATPM) to analyse four sets of
thermal infrared data obtained by IRAS, AKARI, WISE and Subaru/COMICS at different
epochs. In addition, by modeling the thermal lightcurve observed by WISE, we obtain
the rotational phases of each dataset. These rotationally resolved data do not reveal
significant variations of thermal inertia and roughness across the surface, indicating
the surface of Dembowska should be covered by a dusty regolith layer with few rocks or
boulders. Besides, the low thermal inertia of Dembowska show no significant difference
with other asteroids larger than 100 km, indicating the dynamical lives of these large
asteroids are long enough to make the surface to have sufficiently low thermal inertia.
Furthermore, based on the derived surface thermophysical properties, as well as the
known orbital and rotational parameters, we can simulate Dembowska's surface and
subsurface temperature throughout its orbital period. The surface temperature varies
from $\sim40$ K to $\sim220$ K, showing significant seasonal variation, whereas the
subsurface temperature achieves equilibrium temperature about $120\sim160$ K below
$30\sim50$ cm depth.
\end{abstract}

\begin{keywords}
radiation mechanisms: thermal -- minor planets, asteroids: individual:
(349) Dembowska -- infrared: general
\end{keywords}

\section{Introduction}
Asteroid (349) Dembowska was discovered on December 9, 1892. The asteroid
locates in the main belt, orbiting around the Sun with a semi-major axis of
2.92 AU. The orbit of Dembowska is nearly circular, and show a prominent 7:3
resonance with Jupiter.

Dembowska was observed to be a bright asteroid with an absolute magnitude
$H_{\rm v}\approx5.93$, a phase slope parameter $G\approx0.37$ \citep{JPL,MPC}.
Thus Dembowska is believed to be a relatively large asteroid in the main belt.
The effective diameter of Dembowska was derived to be $\sim$ $140~\rm km$
from the STM (Standard thermal model) fitting to IRAS data \citep{Tedesco1989},
indicating an unusually high geometric albedo about $0.38$.

Dembowska is classified to be the unique R-type asteroid, because its near infrared
reflectance spectrum exhibits two strong absorption features at 1 and 2 $\mu m$,
indicating an olivine-pyroxene mixture with little or no metal, where the pyroxene
may be dominantly a calcium-poor and low iron ($\sim F_{s10-30}$) orthopyroxene like
those in ordinary chondrites \citep{Gaffey1993}. Moreover, the Dembowska's spectrum,
reminiscent of (4) Vesta, suggests that the asteroid may have undergone partial melting
or differentiation, making it a good target to study thermal melting and differentiation
history of minor planets, where the accurate size, density, porosity and thermal state
of Dembowska are required to serve as basic constraints on thermal differentiation models.

However, new observations of AKARI and WISE indicate the size of Dembowska could
be much larger, and its albedo may not be so high as the previous reported value.
\citet{Hanus2013} showed the effective diameter of Dembowska derived from data detected
by three space telescopes --- IRAS, AKARI, and WISE, giving $D_{\rm IRAS}=139.8\pm4.3~\rm km$,
$D_{\rm AKARI}=164.7\pm1.8~\rm km$ and $D_{\rm WISE}=216.7\pm7.4~\rm km$, respectively.
The derived sizes from different dataset significantly differ from each other, raising
the question what the accurate size and albedo of Dembowska should be.

We note that the initial result of $D_{\rm IRAS}=139.8\pm4.3~\rm km$ was obtained
by the STM, whereas the update results of $D_{\rm AKARI}=164.7\pm1.8~\rm km$ and
$D_{\rm WISE}=216.7\pm7.4~\rm km$ were derived by the NEATM \citep{Harris1998}. The
various utilized models may lead to the difference in outcomes, while the difference
between $D_{\rm AKARI}$ and $D_{\rm WISE}$ may  mainly come from their different
observation geometries, because WISE observed Dembowska nearly in an equatorial view,
which has a larger cross-sectional area than the south region as observed by AKARI,
according to Dembowska's 3D shape model constructed by \citet{Torppa2003} with the
light-curve inversion method developed by \citet{Kaasalainen2001}. Therefore, to
derive a more reliable and accurate size of the asteroid, we should adopt a more
advanced thermophysical model to combine all available data together rather than
use them separately.

On the other hand, \citet{Torppa2003} derived the rotation period of Dembowska to
be about $4.701\rm~h$, indicating a faster rotation than many large main belt
asteroids, and gave the orientation of rotation axis to be about $\lambda=150^\circ$,
$\beta=+23^\circ$. Furthermore, \citet{Hanus2013} updated the shape model of
Dembowska, which looks like an asymmetrically elongated ellipsoid, and gave the
best-fit spin axis orientation to be $\lambda=322^\circ$, $\beta=+18^\circ$.
The spin axis nearly lies on the orbital plane, which could cause
significant seasonal variation of surface temperature on high local latitudes,
and thus play a role in the thermophysical properties of the surface materials.
In the present work, we implement the radiometric method, where the Advanced
thermophysical modelling (ATPM) algorithm \citep{Rozitis2011, Yu2014, Yu2015, Yu2017}
is used to analyse four independent mid-infrared datasets (IRAS, WISE, AKARI and Subaru)
of (349) Dembowska, where the details of the data source are provided in section 2.1.
With the radiometric method, the size of Dembowska, as well as the surface thermal
inertia, roughness fraction and geometric albedo are well determined in section 3.
In addition, in section 4, based on the derived surface properties, we investigate
the surface and subsurface thermal state of Dembowska, which show significant
seasonal variations.

\section{Radiometric Procedure}
\subsection{Thermal infrared Observations}
In this work, we use the thermal infrared data provided by the IRAS, AKARI
satellite, and the WISE space telescope as well as a new dataset observed
by the Subaru telescope atop Mauna Kea.

The Subaru observations were carried out on UT January 18, 2014, using the
Cooled MIR Camera and Spectrometer \citep[COMICS;][]{Kataza2000} on the 8.2 m
Subaru Telescope. We adopted the N7.8, N8.7, N9.8, N10.3, N11.6, N12.5 continuum
filters in the N-band and the Q18.8 and Q24.5 filters in the Q-band. Immediately
before and/or after the observations of the target, we observed a nearby flux
standard star selected from \citet{Cohen1999}. Data reduction followed the
procedures described in the Subaru Data Reduction CookBook: COMICS, prepared
by Y. Okamoto and the COMICS team.

The IRAS data were obtained from the IMPS Sightings Data Base of VizieR.
The AKARI data were provided by F. Usui (private comm.). The WISE data are
obtained from the WISE archive. We convert the magnitude data to flux with
color corrections (W3:1.0006; W4:0.9833). The derived monochromatic flux
densities for the W3 and W4 band observations have an associated uncertainty
of $\pm10$ percent \citep{Wright2010}.

All these data are utilized in this work to be compared with the theoretical
flux simulated from the Advanced thermophysical Model (ATPM) so as to derive
the possible scale of surface thermophysical properties. We tabulate all the
utilized data in Table \ref{obs}.

\begin{table*}\footnotesize
\centering
\renewcommand\arraystretch{1.0}
\caption{Mid-infrared observations of 349 Dembowska.}
\label{obs}
\begin{tabular}{@{}ccccccccccc@{}}
\hline
 UT  & \multicolumn{4}{c}{Flux (Jy)} & $r_{\rm helio}$ & $\Delta_{\rm obs}$ & $\alpha$ & Observatory\ \\
     & 12.0 ($\mu m$) & 25.0 ($\mu m$) & 65.0 ($\mu m$) & 100.0 ($\mu m$) & (AU) & (AU) & ($^{\circ}$) & Instrument \\
1983-02-17 07:03 & 6.47$\pm$0.64 & 19.13$\pm$2.77 &  8.91$\pm$1.94 & 2.87$\pm$0.58 & 2.809 & 2.480 &-20.32&IRAS \\
1983-02-17 08:46 & 7.47$\pm$0.81 & 20.87$\pm$2.75 & 10.03$\pm$2.19 & 3.10$\pm$0.67 & 2.809 & 2.481 &-20.33&IRAS \\
1983-02-17 10:29 & 6.81$\pm$0.68 & 19.16$\pm$2.46 & 11.71$\pm$2.81 & 2.61$\pm$0.52 & 2.809 & 2.482 &-20.33&IRAS \\
1983-03-02 06:01 & 5.69$\pm$0.65 & 16.88$\pm$2.24 &  9.08$\pm$2.21 & 2.75$\pm$0.55 & 2.820 & 2.666 &-20.56&IRAS \\
1983-03-02 07:44 & 5.92$\pm$0.59 & 16.06$\pm$2.62 &  8.85$\pm$2.14 & 3.61$\pm$0.79 & 2.820 & 2.667 &-20.56&IRAS \\
1983-03-02 09:27 & 7.46$\pm$0.74 & 19.96$\pm$2.68 & 11.56$\pm$2.58 & 3.67$\pm$0.75 & 2.820 & 2.668 &-20.56&IRAS \\
\\
 UT &\multicolumn{2}{c}{Wavelength} &\multicolumn{2}{c}{Flux} & $r_{\rm helio}$ & $\Delta_{\rm obs}$ & $\alpha$ & Observatory\ \\
    &\multicolumn{2}{c}{($\mu m$)} &\multicolumn{2}{c}{(Jy)} & (AU) & (AU) & ($^{\circ}$) & Instrument \\

 2006-05-11 01:52 & \multicolumn{2}{c}{18.0} & \multicolumn{2}{c}{19.08$\pm$1.27} & 2.858 & 2.684 &  20.68 & AKARI   \\
 2006-05-11 03:31 & \multicolumn{2}{c}{18.0} & \multicolumn{2}{c}{19.05$\pm$1.27} & 2.858 & 2.684 &  20.68 & AKARI   \\
 2006-05-11 11:46 & \multicolumn{2}{c}{9.0 } & \multicolumn{2}{c}{ 3.54$\pm$0.21} & 2.858 & 2.679 &  20.69 & AKARI   \\
 2006-05-11 13:25 & \multicolumn{2}{c}{9.0 } & \multicolumn{2}{c}{ 3.58$\pm$0.22} & 2.858 & 2.678 &  20.69 & AKARI   \\
 2006-05-11 15:05 & \multicolumn{2}{c}{9.0 } & \multicolumn{2}{c}{ 3.40$\pm$0.22} & 2.858 & 2.677 &  20.69 & AKARI   \\
 2006-11-11 20:57 & \multicolumn{2}{c}{18.0} & \multicolumn{2}{c}{20.93$\pm$1.40} & 2.719 & 2.541 & -21.34 & AKARI   \\
 2006-11-11 22:37 & \multicolumn{2}{c}{18.0} & \multicolumn{2}{c}{21.43$\pm$1.43} & 2.719 & 2.541 & -21.34 & AKARI   \\
 2006-11-12 00:16 & \multicolumn{2}{c}{18.0} & \multicolumn{2}{c}{21.17$\pm$1.41} & 2.719 & 2.543 & -21.34 & AKARI   \\
\\
 UT  & \multicolumn{4}{c}{Flux (Jy)} & $r_{\rm helio}$ & $\Delta_{\rm obs}$ & $\alpha$ & Observatory\ \\
     & \multicolumn{2}{c}{11.0 ($\mu m$)} &\multicolumn{2}{c}{22.0 ($\mu m$)}& (AU) & (AU) & ($^{\circ}$) & Instrument \\

 2010-02-14 12:45 & \multicolumn{2}{c}{2.39$\pm$0.24} &\multicolumn{2}{c}{ 7.78$\pm$0.78} & 3.170 & 3.010 &  18.15 & WISE  \\
 2010-02-14 12:46 & \multicolumn{2}{c}{2.39$\pm$0.24} &\multicolumn{2}{c}{ 7.78$\pm$0.78} & 3.170 & 3.010 &  18.15 & WISE  \\
 2010-02-14 22:17 & \multicolumn{2}{c}{2.54$\pm$0.25} &\multicolumn{2}{c}{ 8.47$\pm$0.85} & 3.170 & 3.004 &  18.15 & WISE  \\
 2010-02-14 23:52 & \multicolumn{2}{c}{2.60$\pm$0.26} &\multicolumn{2}{c}{ 8.29$\pm$0.82} & 3.169 & 3.003 &  18.15 & WISE  \\
 2010-02-15 01:28 & \multicolumn{2}{c}{3.65$\pm$0.36} &\multicolumn{2}{c}{11.07$\pm$1.11} & 3.169 & 3.002 &  18.15 & WISE  \\
 2010-02-15 03:03 & \multicolumn{2}{c}{2.65$\pm$0.26} &\multicolumn{2}{c}{ 8.48$\pm$0.84} & 3.169 & 3.001 &  18.15 & WISE  \\
 2010-02-15 12:35 & \multicolumn{2}{c}{3.09$\pm$0.31} &\multicolumn{2}{c}{ 9.81$\pm$0.98} & 3.169 & 2.996 &  18.15 & WISE  \\
 2010-02-15 15:45 & \multicolumn{2}{c}{3.44$\pm$0.34} &\multicolumn{2}{c}{ 9.95$\pm$0.99} & 3.169 & 2.994 &  18.15 & WISE  \\
 2010-08-04 02:37 & \multicolumn{2}{c}{3.37$\pm$0.33} &\multicolumn{2}{c}{10.82$\pm$1.08} & 3.090 & 2.822 & -19.07 & WISE  \\
 2010-08-04 05:47 & \multicolumn{2}{c}{3.04$\pm$0.30} &\multicolumn{2}{c}{ 9.94$\pm$0.99} & 3.090 & 2.823 & -19.07 & WISE  \\
 2010-08-04 08:58 & \multicolumn{2}{c}{4.65$\pm$0.46} &\multicolumn{2}{c}{12.64$\pm$1.26} & 3.090 & 2.825 & -19.08 & WISE  \\
 2010-08-04 12:08 & \multicolumn{2}{c}{3.11$\pm$0.31} &\multicolumn{2}{c}{ 9.78$\pm$0.98} & 3.090 & 2.827 & -19.08 & WISE  \\
 2010-08-04 13:43 & \multicolumn{2}{c}{4.45$\pm$0.44} &\multicolumn{2}{c}{12.95$\pm$1.29} & 3.090 & 2.828 & -19.08 & WISE  \\
 2010-08-04 15:19 & \multicolumn{2}{c}{2.82$\pm$0.28} &\multicolumn{2}{c}{ 9.46$\pm$0.94} & 3.089 & 2.829 & -19.08 & WISE  \\
 2010-08-04 16:54 & \multicolumn{2}{c}{3.49$\pm$0.34} &\multicolumn{2}{c}{10.16$\pm$1.01} & 3.089 & 2.830 & -19.08 & WISE  \\
 2010-08-04 23:15 & \multicolumn{2}{c}{4.07$\pm$0.41} &\multicolumn{2}{c}{11.90$\pm$1.19} & 3.089 & 2.833 & -19.09 & WISE  \\
 2010-08-05 02:25 & \multicolumn{2}{c}{2.97$\pm$0.30} &\multicolumn{2}{c}{ 9.40$\pm$0.94} & 3.089 & 2.835 & -19.09 & WISE  \\
 2010-08-05 05:36 & \multicolumn{2}{c}{2.88$\pm$0.29} &\multicolumn{2}{c}{ 9.69$\pm$0.97} & 3.089 & 2.836 & -19.10 & WISE  \\
 \\
 UT &\multicolumn{2}{c}{Wavelength} &\multicolumn{2}{c}{Flux} & $r_{\rm helio}$ & $\Delta_{\rm obs}$ & $\alpha$ & Observatory\ \\
    &\multicolumn{2}{c}{($\mu m$)} &\multicolumn{2}{c}{(Jy)} & (AU) & (AU) & ($^{\circ}$) &  \\
 2014-01-18 12:39 & \multicolumn{2}{c}{~~7.8} & \multicolumn{2}{c}{ 2.85$\pm$0.45} & 3.086 & 2.337 &  13.63 & Subaru  \\
 2014-01-18 12:43 & \multicolumn{2}{c}{~~8.7} & \multicolumn{2}{c}{ 3.84$\pm$0.47} & 3.086 & 2.337 &  13.63 & Subaru  \\
 2014-01-18 12:47 & \multicolumn{2}{c}{~~9.8} & \multicolumn{2}{c}{ 6.02$\pm$0.72} & 3.086 & 2.337 &  13.63 & Subaru  \\
 2014-01-18 12:51 & \multicolumn{2}{c}{10.3} & \multicolumn{2}{c}{ 8.10$\pm$0.94} & 3.086 & 2.337 &  13.63 & Subaru  \\
 2014-01-18 12:54 & \multicolumn{2}{c}{11.6} & \multicolumn{2}{c}{11.11$\pm$1.26} & 3.086 & 2.337 &  13.63 & Subaru  \\
 2014-01-18 12:58 & \multicolumn{2}{c}{12.5} & \multicolumn{2}{c}{12.44$\pm$1.42} & 3.086 & 2.337 &  13.63 & Subaru  \\
 2014-01-18 13:04 & \multicolumn{2}{c}{18.7} & \multicolumn{2}{c}{21.74$\pm$2.71} & 3.086 & 2.337 &  13.63 & Subaru  \\
 2014-01-18 13:10 & \multicolumn{2}{c}{24.5} & \multicolumn{2}{c}{25.82$\pm$5.57} & 3.086 & 2.337 &  13.63 & Subaru  \\

\hline
\end{tabular}
\end{table*}

\subsection{Advanced thermophysical model}
The Advanced thermophysical model reproduces the thermal state and thermal emission
of an asteroid by solving 1D thermal conduction in consideration of roughness, where
the asteroid is described by a polyhedron composed of $N$ triangle facets, and
the roughness is modelled by a fractional coverage of hemispherical macroscopic craters,
symbolized by $f_{\rm r}$ ($0\leq f_{\rm r}\leq 1$), while the remaining fraction,
$1-f_{\rm r}$, represents smooth flat surface ($f_{\rm r}=0$ means the whole surface
is smooth flat).

For such rough surface facets, the conservation of energy leads to an instant heat
balance between sunlight, multiple-scattered sunlight, thermal emission, thermal-radiated
fluxes from other facets and heat conduction. If the asteroid keeps a periodical rotation,
the temperature $T_{i}$ and thermal emission $B_{i}$ of facet $i$ will change periodically
as well. Therefore, we can build numerical codes to simulate $T_{i}$ and $B_{i}$ at any
rotation phase for the asteroid. For a given observation epoch, ATPM can reproduce a
theoretical profile to each observation flux as:
\begin{equation}
F_{\rm model}(\lambda)=\sum^{N}_{i=1}\epsilon(\lambda)\pi B(\lambda, T_{i})S(i)f(i)~,
\label{fmodel}
\end{equation}
where $\epsilon(\lambda)$ is the monochromatic emissivity at wavelength $\lambda$,
$S(i)$ is the area of facet $i$, $f(i)$ is the view factor of facet $i$ to the telescope
\begin{equation}
f(i)=v_i\frac{\vec{n}_{i}\cdot\vec{n}_{\rm obs}}{\pi\Delta^{2}},~
\end{equation}
$v_{\rm i}$ indicates visible fraction of facet $i$, and $B(\lambda, T_{i})$ is
the Planck intensity function:
\begin{equation}
B(\lambda, T_{i})=\frac{2hc^{2}}{\lambda^{5}}\frac{1}{ \exp\big(\frac{hc}{\lambda kT_{i}}\big)-1 }~.
\end{equation}
Thus the calculated $F_{\rm model}$ can be compared with the thermal
infrared fluxes summarized in Table \ref{obs} in the fitting process.

\subsection{Fitting Procedure}
In order to reproduce thermal state and thermal emission of an asteroid via
the ATPM procedure, several input physical parameters are needed, including
the 3D shape model, effective diameter $D_{\rm eff}$, bond albedo, and the
so-called thermal parameter
\begin{equation}
\Phi=\frac{\Gamma\sqrt{\omega}}{\varepsilon\sigma T^{3}_{\rm eff}}~,
\end{equation}
where $\omega$ is the rotation frequency, $\Gamma$ is the thermal inertia,
$\varepsilon$ is the averaged thermal emissivity over the entire emission spectrum, and
\[T_{\rm eff}=\left[\frac{(1-A_{\rm B})F_{\odot}}{\varepsilon\sigma d_{\rm \odot}^2}\right]^{1/4}~,\]
is the effective temperature. The rotation frequency $\omega$ can be determined
from light curves, while thermal inertia $\Gamma$ is the parameter of interest
which would be treated as free parameter in the fitting procedure.

\begin{figure}
\centering
\includegraphics[scale=0.63]{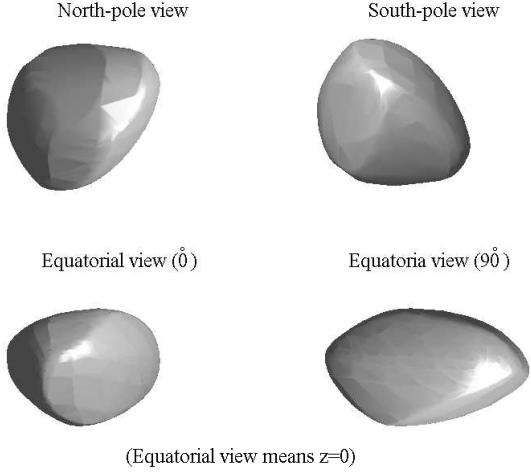}
\caption{The shape model of 349 Dembowska downloaded from DAMIT
 (Database of Asteroid Models from Inversion Techniques).}
\label{shape}
\end{figure}

Figure \ref{shape} shows the 3D shape model of 349 Dembowska published in the
Database of Asteroid Models from Inversion Techniques (DAMIT). This shape model
is updated by \citet{Hanus2013} according to the light-curve inversion method
developed by \citet{Kaasalainen2001}. We utilize this shape model in our
thermophysical modelling procedure.

According to \citet{Fowler1992}, an asteroid's effective diameter $D_{\rm eff}$,
defined by the diameter of a sphere with a identical volume to that of the shape
model, can be related to its geometric albedo $p_{v}$ and absolute visual
magnitude $H_{v}$ via:
\begin{equation}
D_{\rm eff}=\frac{1329\times 10^{-H_{v}/5}}{\sqrt{p_{v}}}~(\rm km) ~.
\label{Deff}
\end{equation}
In addition, the geometric albedo $p_{v}$ is related to the effective
Bond albedo $A_{\rm eff,B}$ by
\begin{equation}
A_{\rm eff,B}=p_{v}q_{\rm ph}~,
\label{aeffpv}
\end{equation}
where $q_{\rm ph}$ is the phase integral that can be approximated by
\begin{equation}
q_{\rm ph}=0.290+0.684G~,
\label{qph}
\end{equation}
in which $G$ is the slope parameter in the $H, G$ magnitude system of
\citet{Bowell}. We obtain $H_{\rm v}=5.93, G=0.37$ from the JPL Website
($https://ssd.jpl.nasa.gov/sbdb.cgi\#top$).

On the other hand, the asteroid's effective Bond albedo is the averaged
result of both the albedo of smooth and rough surface, which can be expressed
as the following relationship:
\begin{equation}
A_{\rm eff,B}=(1-f_{\rm r})A_{B}+f_{\rm r}\frac{A_{B}}{2-A_{B}}~,
\label{abfr}
\end{equation}
where $A_{B}$ is the Bond albedo of smooth lambertian surface. Thus an input
roughness fraction $f_{\rm r}$ and geometric albedo $p_{\rm v}$ can lead to
an unique Bond albedo $A_{B}$ and effective diameter $D_{\rm eff}$ to be used
to fit the observations.

Then we actually have three free parameters --- thermal inertia, roughness fraction,
and geometric albedo (or effective diameter) that would be extensively investigated
in the fitting process. We use the so-called reduced $\chi^{2}_{\rm r}$ defined as
\begin{equation}
\chi^{2}_{\rm r}=\frac{1}{n-3}\sum^{n}_{i=1}
\Big[\frac{F_{\rm model}(\lambda_i,\Gamma,f_{\rm r},p_{\rm v})
    -F_{\rm obs}(\lambda_{i})}{\sigma_{\lambda_{i}}}\Big]^{2}~,
\label{l2}
\end{equation}
to assess the fitting degree of our model with respect to the observations.
Other parameters are listed in Table \ref{phpa}.
\begin{table}\footnotesize
 \centering
 \renewcommand\arraystretch{1.3}
 \caption{Assumed physical parameters used in ATPM.}
 \label{phpa}
 \begin{tabular}{@{}lcc@{}}
 \hline
 Property & Value & References \\
 \hline
 Number of vertices  &     1022               & \citep{Hanus2013}  \\
 Number of facets    &     2040               & \citep{Hanus2013}   \\
 Shape (a:b:c)       & 1.4165:1.2569:1        & \citep{Hanus2013}   \\
 Spin axis           & ($322.0^\circ$,$18.0^{\circ}$) & \citep{Hanus2013} \\
 Spin period         &     4.701  h           & \citep{Torppa2003}  \\
 Absolute magnitude  &     5.93               & \citep{JPL,MPC} \\
 Slope parameter     &     0.37               & \citep{JPL,MPC} \\
 Emissivity $\varepsilon$  &     0.9    & Assumed \\
 Emissivity $\epsilon(\lambda)$ & 0.9 & Assumed \\
 \hline
\end{tabular}
\end{table}

\section{Analysis and Results}
\subsection{Fitting with rotationally averaged flux}
Due to the uncertainties of rotation phases at different observation epochs, we
choose rotationally averaged model flux $F_{\rm model}$ to fit the observations.
Thus the investigated thermophysical parameters would be averaged profiles across
the whole surface. Table \ref{fitchi2} lists the reduced $\chi^{2}_{\rm r}$ derived
from each input parameters, where we scan the thermal inertia in the range of
$0\sim50\rm~Jm^{-2}s^{-0.5}K^{-1}$ and roughness fraction $0\sim1.0$, while for
each pair of thermal inertia and roughness fraction, the geometric albedo giving
the minimum reduced $\chi^{2}_{\rm r}$ is found.
\begin{table*}
\vskip 44pt
\renewcommand\arraystretch{1.0}
\centering
\caption{ATPM fitting results to the observations.}
\label{fitchi2}
\begin{tabular}{@{}p{0.5cm}C{0.5cm}C{0.5cm}C{0.5cm}C{0.5cm}C{0.5cm}C{0.5cm}C{0.5cm}C{0.5cm}
C{0.5cm}C{0.5cm}C{0.5cm}C{0.5cm}C{0.5cm}C{0.5cm}C{0.5cm}C{0.5cm}C{0.5cm}C{0.5cm}@{}}
\hline
 Roughness &\multicolumn{18}{c}{Thermal inertia $\Gamma$ ($\rm~Jm^{-2}s^{-0.5}K^{-1}$)} \\
 fraction & \multicolumn{2}{c}{5}  & \multicolumn{2}{c}{10} & \multicolumn{2}{c}{15}
          & \multicolumn{2}{c}{20} & \multicolumn{2}{c}{25} & \multicolumn{2}{c}{30}
          & \multicolumn{2}{c}{35} & \multicolumn{2}{c}{40} & \multicolumn{2}{c}{45} \\
$f_{\rm R}$ & $p_{\rm v}$ & $\chi^{2}_{\rm r}$ & $p_{\rm v}$ & $\chi^{2}_{\rm r}$
            & $p_{\rm v}$ & $\chi^{2}_{\rm r}$ & $p_{\rm v}$ & $\chi^{2}_{\rm r}$
            & $p_{\rm v}$ & $\chi^{2}_{\rm r}$ & $p_{\rm v}$ & $\chi^{2}_{\rm r}$
            & $p_{\rm v}$ & $\chi^{2}_{\rm r}$ & $p_{\rm v}$ & $\chi^{2}_{\rm r}$
            & $p_{\rm v}$ & $\chi^{2}_{\rm r}$  \\
\hline
0.00&0.325&2.572&0.310&2.199&0.297&1.989&0.287&1.908&0.278&1.931&0.271&2.035&0.264&2.201&0.258&2.415&0.254&2.664 \\
0.05&0.328&2.604&0.315&2.201&0.304&1.974&0.295&1.885&0.287&1.906&0.280&2.012&0.274&2.183&0.269&2.404&0.264&2.661 \\
0.10&0.333&2.642&0.319&2.209&0.308&1.966&0.298&1.868&0.290&1.887&0.283&1.995&0.277&2.171&0.272&2.399&0.267&2.664 \\
0.15&0.338&2.685&0.324&2.224&0.312&1.963&0.302&1.857&0.294&1.874&0.286&1.984&0.280&2.165&0.275&2.399&0.270&2.672 \\
0.20&0.343&2.732&0.328&2.243&0.316&1.966&0.306&1.852&0.297&1.865&0.290&1.978&0.283&2.163&0.278&2.404&0.273&2.685 \\
0.25&0.347&2.783&0.332&2.267&0.320&1.974&0.309 &1.851 
&0.300&1.862&0.293&1.976&0.286&2.166&0.280&2.413&0.275&2.701 \\
0.30&0.352&2.837&0.336&2.294&0.323&1.985&0.312&1.854&0.303&1.863&0.295&1.978&0.289&2.173&0.283&2.426&0.278&2.721 \\
0.35&0.356&2.893&0.340&2.324&0.327&2.000&0.315&1.861&0.306&1.867&0.298&1.984&0.291&2.183&0.285&2.442&0.280&2.745 \\
0.40&0.360&2.951&0.343&2.357&0.330&2.017&0.318&1.871&0.309&1.875&0.301&1.994&0.294&2.197&0.288&2.462&0.282&2.771 \\
0.45&0.364&3.011&0.347&2.393&0.333&2.038&0.321&1.884&0.311&1.886&0.303&2.006&0.296&2.213&0.290&2.484&0.284&2.801 \\
0.50&0.367&3.072&0.350&2.430&0.336&2.061&0.324&1.900&0.314&1.899&0.305&2.021&0.298&2.232&0.292&2.509&0.286&2.832 \\
0.55&0.370&3.135&0.353&2.469&0.338&2.086&0.326&1.918&0.316&1.915&0.307&2.039&0.300&2.254&0.294&2.536&0.288&2.866 \\
0.60&0.373&3.197&0.355&2.509&0.341&2.113&0.328&1.938&0.318&1.933&0.309&2.058&0.302&2.277&0.295&2.565&0.290&2.901 \\
0.65&0.376&3.260&0.358&2.551&0.343&2.141&0.330&1.960&0.320&1.953&0.311&2.079&0.304&2.303&0.297&2.596&0.291&2.939 \\
0.70&0.379&3.324&0.360&2.593&0.345&2.171&0.332&1.983&0.322&1.975&0.313&2.103&0.305&2.330&0.298&2.628&0.293&2.977 \\
0.75&0.381&3.387&0.362&2.637&0.347&2.202&0.334&2.008&0.323&1.998&0.314&2.127&0.306&2.358&0.300&2.662&0.294&3.017 \\
0.80&0.383&3.451&0.364&2.681&0.348&2.234&0.335&2.034&0.324&2.022&0.315&2.153&0.307&2.388&0.301&2.697&0.295&3.059 \\
0.85&0.385&3.514&0.365&2.725&0.350&2.268&0.336&2.061&0.326&2.048&0.316&2.181&0.308&2.419&0.302&2.733&0.296&3.101 \\
0.90&0.386&3.577&0.366&2.770&0.351&2.301&0.337&2.089&0.326&2.075&0.317&2.209&0.309&2.451&0.302&2.771&0.297&3.145 \\
0.95&0.387&3.640&0.367&2.815&0.352&2.335&0.338&2.118&0.327&2.102&0.318&2.238&0.310&2.484&0.303&2.809&0.297&3.188 \\
1.00&0.424&3.702&0.393&2.860&0.369&2.370&0.350&2.147&0.334&2.130&0.321&2.269&0.310&2.518&0.301&2.848&0.293&3.233 \\
\hline
\end{tabular}
\end{table*}

According to Table \ref{fitchi2}, we can see that a low thermal inertia
between $20\sim30\rm~Jm^{-2}s^{-0.5}K^{-1}$ tends to fit better to the
observations; the minimum reduced $\chi^{2}_{\rm r}$ arises around the case
of $p_{\rm v}=0.309$, $f_{\rm r}=0.25$, and $\Gamma=20$$\rm~Jm^{-2}s^{-0.5}K^{-1}$,
which can be adopted as the best solution for the geometric albedo, thermal
inertia and roughness fraction of Dembowska.

\begin{figure}
\includegraphics[scale=0.63]{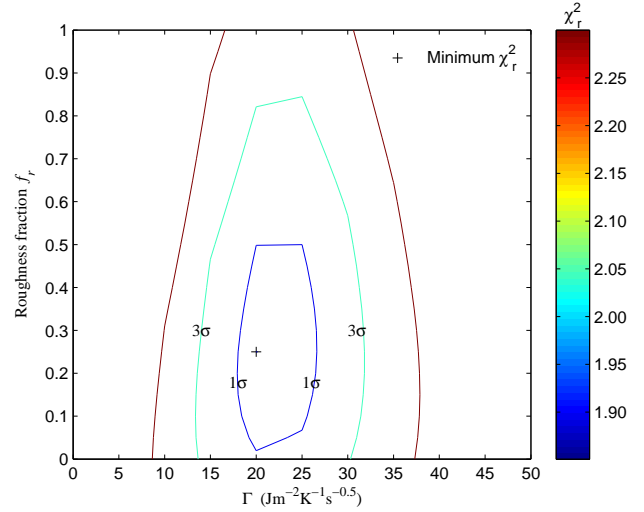}
  \centering
  \caption{$\chi^{2}_{\rm r}$ ($f_{\rm r}$, $\Gamma$) contour according to Table \ref{fitchi2}.
  The color (from blue to red) means the increase of profile of $\chi^{2}_{\rm r}$.
  The 1$\sigma$ boundary corresponds to $\Delta\chi^2_{\rm r}=3.52/(76-3)=0.0482$, while
  the 3$\sigma$ boundary corresponds to $\Delta\chi^2_{\rm r}=14.2/(76-3)=0.1945$
  \citep{Press2007}.
  }\label{contour}
\end{figure}

Figure \ref{contour} shows the contour of $\chi^2_{\rm r}(\Gamma,f_{\rm r})$
based on the results listed in Table \ref{fitchi2}, where the values of
$\chi^{2}_{\rm r}$ are represented by colour and the variation of ColorBar
from blue to red means the increase of $\chi^{2}_{\rm r}$. The black '+'
shows the location of minimum $\chi^{2}_{\rm r}$ in the ($\Gamma$, $f_{\rm r}$)
parameter space. The blue curve labeled by 1$\sigma$ corresponding to
$\Delta\chi^{2}_{\rm r}=3.52/(76-3)=0.0482$ from the minimum
$\chi^{2}_{\rm r}$, which constrains the range of free parameters
$\Gamma=20^{+6}_{-3}\rm~Jm^{-2}s^{-0.5}K^{-1}$, $f_{\rm r}=0\sim0.5$,
with probability of $68.3\%$, while the cyan curve labeled by 3$\sigma$
refers to $\Delta\chi^{2}_{\rm r}=14.2/(76-3)=0.1945$, giving the range
of free parameters $\Gamma=20^{+12}_{-7}\rm~Jm^{-2}s^{-0.5}K^{-1}$,
$f_{\rm r}=0\sim0.85$ with probability of $99.73\%$ \citep{Press2007}.

\begin{figure}
\includegraphics[scale=0.63]{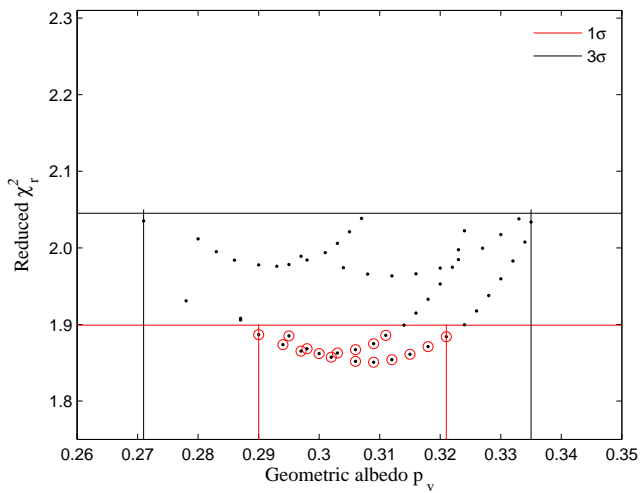}
  \centering
  \caption{$p_{\rm v}\sim\chi^2_{\rm reduced}$ profiles fit to the observations
  in consideration of all $f_{\rm r}$ and $\Gamma$ in the derived 1$\sigma$ and
  3$\sigma$ range.
  }\label{galbedo}
\end{figure}

In consideration of the above derived 1$\sigma$ and 3$\sigma$ range of
thermal inertia $\Gamma$ and roughness fraction $f_{\rm r}$, the corresponding
geometric albedo $p_{\rm v}$ and $\chi^2_{\rm r}$ are selected out, yielding the
$p_{\rm v}\sim\chi^2_{\rm r}$ relation showed in Figure \ref{galbedo}. Then the
1$\sigma$ scale of geometric albedo can be constrained to be
$p_{\rm v}=0.309^{+0.012}_{-0.019}$, while the 3$\sigma$ scale is
$p_{\rm v}=0.309^{+0.026}_{-0.038}$.

\begin{figure}
\includegraphics[scale=0.6]{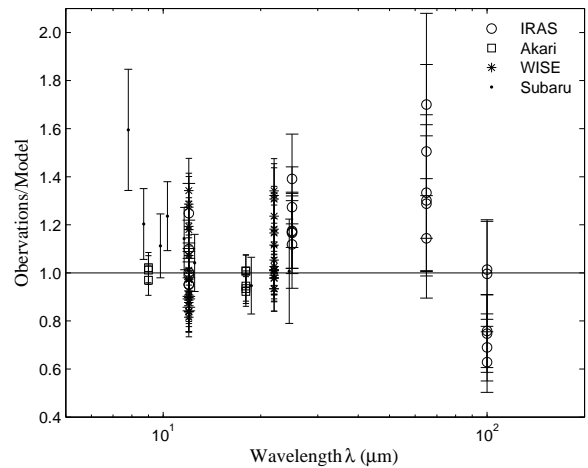}
\centering
\caption{The observation/model ratios as a function of wavelength
  for $\Gamma=20\rm~Jm^{-2}s^{-0.5}K^{-1}$, $f_{\rm r}=0.25$, $p_{\rm v}=0.309$
  and $D_{\rm eff}=155.8\rm~km$.
  }\label{nspectra}
\end{figure}

\begin{figure}
\includegraphics[scale=0.6]{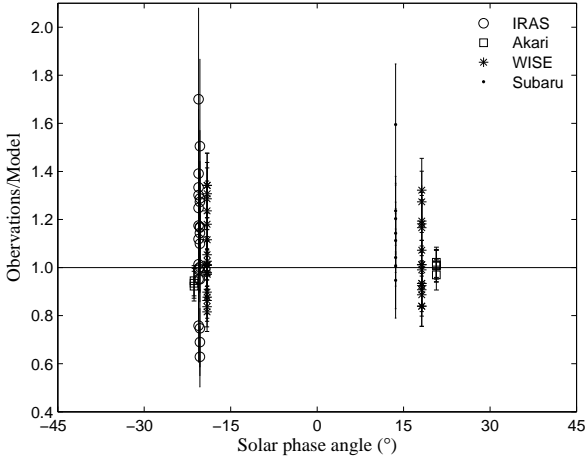}
\centering
\caption{The observation/model ratios as a function of solar phase angle
  for $\Gamma=20\rm~Jm^{-2}s^{-0.5}K^{-1}$, $f_{\rm r}=0.25$, $p_{\rm v}=0.309$
  and $D_{\rm eff}=155.8\rm~km$.
  }\label{nsolarph}
\end{figure}

To verify the the reliability of the above fitting procedure and derived outcomes,
we employ the ratio of 'observation/model' to examine how these theoretical model
results match the observations at various observation wavelengths and observation
geometries (see Figure \ref{nspectra} and \ref{nsolarph}), because these factors
are the basic variables of the observations.

In Figure \ref{nspectra}, the observation/Model ratios are shown at each
observational wavelength for $\Gamma=20\rm~Jm^{-2}s^{-0.5}K^{-1}$, $f_{\rm r}=0.25$,
$p_{\rm v}=0.309$ and $D_{\rm eff}=155.8\rm~km$. The ratios are evenly distributed
around 1.0 without significant wavelength dependent features, despite the ratio
at $7.8\rm~\mu m$ moves relatively farther from unity. In Figure \ref{nsolarph},
we show how the model results match observations at each observational solar phase
angle, where the ratios are all nearly symmetrical distributed around 1.0 and no
phase-angle dependent features exit as well. Thus the relatively large deviation
for the $7.8\rm~\mu m$ data observed by Subaru may be caused by large observation
uncertainty. We note that ground-based observations are vulnerable to telluric
absorptions. The atmosphere at $7.8\rm~\mu m$ is only about $60-80\%$ transparent,
therefore it is expected that the uncertainty of the flux measurement at this band
is relatively higher than other wavelengths. The telluric contamination explains
the relatively large deviation of the $7.8\rm~\mu m$ data from the model.
Nevertheless, the fitting procedure is reliable, but more accurate data especially
that observed at low phase angle and at wavelength around Wien peak are needed to
examine our results. We summarize all the derived results from the above
thermophysical modelling process in Table \ref{results}.
\begin{table}\footnotesize
 \centering
 \renewcommand\arraystretch{2}
 \caption{Derived Results from rotationally averaged fitting.}
 \label{results}
 \begin{tabular}{@{}lcc@{}}
 \hline
  Properties & $1\sigma$ value & $3\sigma$ value \\
 \hline
  Thermal inertia $\Gamma$($\rm~Jm^{-2}s^{-0.5}K^{-1}$)  & $20^{+6}_{-3}$ & $20^{+12}_{-7}$ \\
  Roughness fraction $f_{\rm r}$   & $0\sim0.5$  & $0\sim0.85$ \\
  Geometric albedo $p_{\rm v}$   & $0.309^{+0.012}_{-0.019}$ & $0.309^{+0.026}_{-0.038}$ \\
  Effective diameter $D_{\rm eff}$ (km) & $155.8^{+5.0}_{-3.0}$ & $155.8^{+7.5}_{-6.2}$  \\
 \hline
\end{tabular}
\end{table}

\subsection{Fitting with thermal light curve}
Since the IRAS and WISE data do not cover an entire rotation period and
were observed at various solar phase angles, they cannot be used to generate
thermal light curves directly. However, Dembowska is a large and well-observed
asteroid with known orbital and rotational parameters, thus in principle, we
could derive the rotational phase of each observation data with respect to a
defined local body-fixed coordinate system if we know the observed rotational
phase at a particular epoch, then these data can be used to create thermal
light curves.

We use the published 3D shape model of Dembowska showed in Figure \ref{shape}
to define the local body-fixed coordinate system, where the z-axis is chosen
to be the rotation axis, and "zero" rotational phase is chosen to be the
"Equatorial view ($0^\circ$)" in Figure \ref{shape}. Moreover, if we define
the view angle of one observation with respect to the body-fixed coordinate
system to be $(\varphi,\theta)$, where $\varphi$ stands for local longitude,
and $\theta$ means local latitude, then the rotational phase $ph$ of this
observation can be related to the local longitude $\varphi$ via
\begin{equation}
ph=1-\varphi/(2\pi).
\end{equation}

\begin{figure}
\includegraphics[scale=0.63]{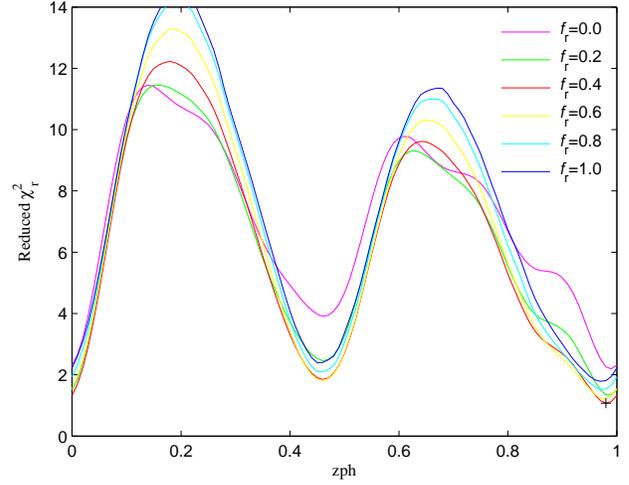}
  \centering
  \caption{Horizontal axis: $zph$ represents the observed rotational phase of Dembowska
  at epoch 2010-02-14 12:45. Vertical axis: obtained reduced $\chi^2_{\rm r}$ with different
  input $zph$ and roughness fraction $f_{\rm r}$. The curves in different color stand for
  different $f_{\rm r}$ ranging from $0.0\sim1.0$.
  }\label{zph}
\end{figure}

If assuming the rotational phase at epoch 2010-02-14 12:45 to be $zph$, then
all the rotational phases of other data could be derived in consideration of
the observation time and geometry. However, the observation time of the IRAS
data deviate so long away from the reference epoch 2010-02-14 12:45 that even
tiny uncertainty of rotation period can accumulate significant estimation
errors of the rotational phases. Thus we only fit the rotationally resolved
flux data of WISE so as to find out which $zph$ can fit the data best. Of course,
other parameters mentioned above are necessary. Since the surface thermal inertia,
albedo and size are well constrained within 3$\sigma$ by the above rotationally
averaged fitting, we could use the derived best-fit profiles as definite parameters.
But for the roughness fraction, the uncertainty is relatively larger. Therefore we
use $f_{\rm r}$ and $zph$ together as free parameters to fit the WISE data, and
the results are showed in Figure \ref{zph}.

\begin{figure}
\includegraphics[scale=0.63]{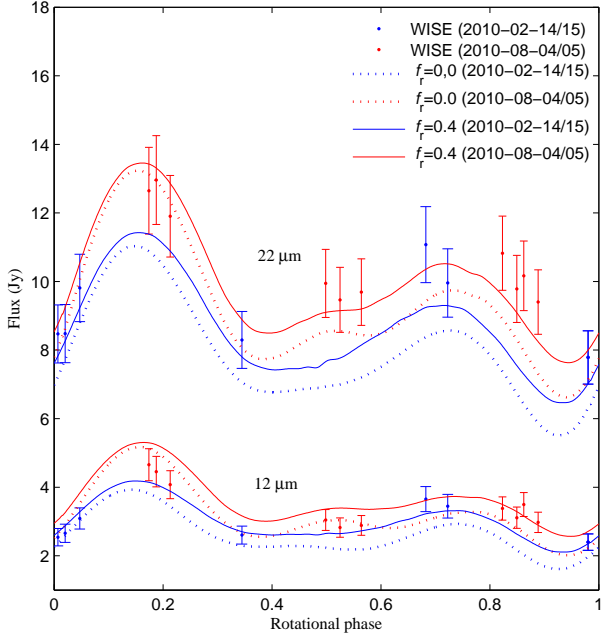}
  \centering
  \caption{Thermal light curve of Dembowska. The blue data corresponds to epoch 2010-02-14 12:45,
  the red data corresponds to epoch 2010-08-04 02:37; the colorized curves are modeled with
  $\Gamma=20\rm~Jm^{-2}s^{-0.5}K^{-1}$, $p_{\rm v}=0.309$, $D_{\rm eff}=155.8\rm~km$, while
  the dashed curves refer to modeled curves with input roughness fraction $f_{\rm r}=0.0$,
  and the solid curves refer to $f_{\rm r}=0.4$.
  }\label{thlc}
\end{figure}

Figure \ref{zph} shows how different $zph$ and roughness fraction $f_{\rm r}$ match
the observations, where $zph=0.98$ and $f_{\rm r}\sim0.4$ (shown by the black '+' in
Figure \ref{zph}) seem to achieve best degree of fitting. Thus we adopt $zph=0.98$
as the rotational phase of Dembowska at epoch 2010-02-14 12:45, and use it to derive
the rotational phases of other data.
With the derived rotational phases, we are able to convert the WISE data into one
rotation period, and make comparisons with the modeled thermal light curve, as
showed in Figure \ref{thlc}, where the colorized curves are the modeled thermal
light curves with the above best-fit results of $\Gamma=20\rm~Jm^{-2}s^{-0.5}K^{-1}$,
$p_{\rm v}=0.309$, $D_{\rm eff}=155.8\rm~km$, and two different roughness fraction
cases $f_{\rm r}=0.0,0.4$. The difference between the blue and red curves are
caused by different heliocentric and observation distance at each epoch.
Figure \ref{thlc} shows that the modeled curves with higher roughness
fraction $f_{\rm r}\approx0.4$ tends to better match the observations.

\begin{figure}
\includegraphics[scale=0.63]{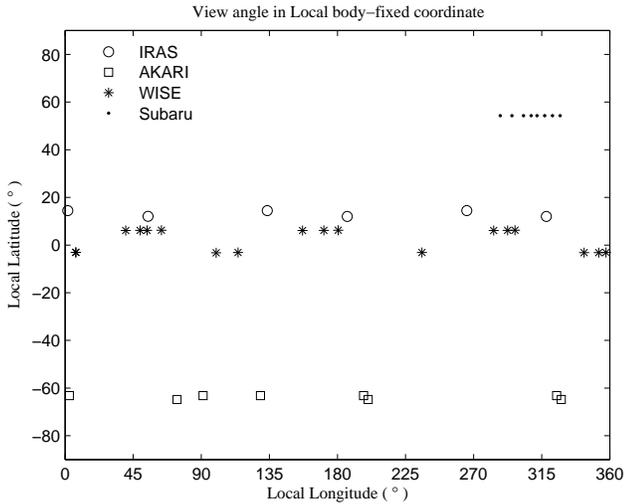}
  \centering
  \caption{View angles of each observation with respect to the defined local body-fixed
  coordinate system.
  }\label{viewangle}
\end{figure}

It should be noticed here that the observations differ from each other not only
in rotational phase, but also in view angle. With the above determined rotational
phases, we can derive the exact view angle of each observation with respect to the
defined local body-fixed coordinate system, and show them in Figure \ref{viewangle},
which exhibits that IRAS and WISE observed Dembowska nearly in equatorial region,
whereas AKARI observed south region and Subaru observed north region.

\begin{figure}
\includegraphics[scale=0.63]{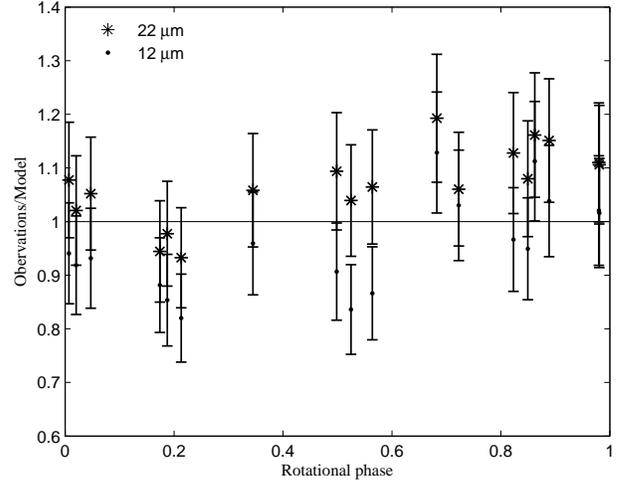}
  \centering
  \caption{The observation/model ratios as a function of rotational phase in equatorial view
  for $f_{\rm r}=0.4$, $\Gamma=20\rm~Jm^{-2}s^{-0.5}K^{-1}$, $p_{\rm v}=0.309$
  and $D_{\rm eff}=155.8\rm~km$.
  }\label{nwise}
\end{figure}

\begin{figure}
\includegraphics[scale=0.63]{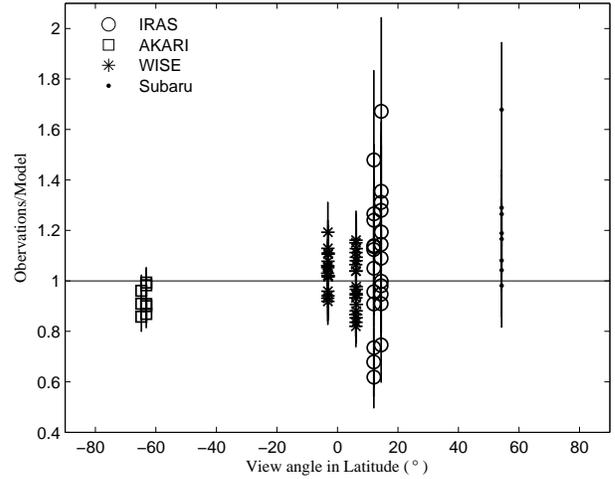}
  \centering
  \caption{The observation/model ratios as a function of view angle in local latitude
  for $f_{\rm r}=0.4$, $\Gamma=20\rm~Jm^{-2}s^{-0.5}K^{-1}$, $p_{\rm v}=0.309$
  and $D_{\rm eff}=155.8\rm~km$.
  }\label{nlats}
\end{figure}

We could investigate whether heterogeneous feature of surface thermophysical
properties appear along local longitude by checking how the observation/Model
ratios of WISE data vary with rotational phase in Figure \ref{nwise}, where
the modeled fluxes are obtained by ATPM with the above determined best-fit
parameters $f_{\rm r}=0.4$, $\Gamma=20\rm~Jm^{-2}s^{-0.5}K^{-1}$, $p_{\rm v}=0.309$
and $D_{\rm eff}=155.8\rm~km$. The ratios are distributed nearly around 1.0 without
significant rotational phase dependent features, despite a slight increasing
tendency from rotational phase $0.2$ to $0.8$, which may indicate heterogeneous
surface properties between the West and East part of Dembowska. But this kind of
heterogeneous signal is rather weak to further infer variation of surface properties
along local longitude in consideration of observation uncertainties. On the other hand,
Figure \ref{nlats} shows the observation/Model ratios corresponding to different view
latitudes, where the ratios tend to be $<1.0$ for south region but $>1.0$ for north
region, indicating a slight heterogeneous feature between the south and north region
of Dembowska. However, the heterogeneous signal is also very week due to the relatively
large observational uncertainties. Thus we may surmise that no significant large variation
of surface thermophysical characteristics appear over the surface of Dembowska.

\section{Surface Thermal State}
In this section, we use the above derived surface thermophysical properties
to further investigate the surface and subsurface thermal environment of Dembowska
based on its present rotational and orbital motion state.
In consideration of the large axial tilt between the rotational axis and orbital
axis, it can be imagined that interesting seasonal variation of temperature would
appear on the surface of Dembowska, which may have potential influence both on
the thermophysical and spectral properties of the surface materials.

Temperature variation can be caused by both orbital motion and rotation. Thus,
to figure out the seasonal variation of temperature, we have to remove the diurnal
effect caused by rotation, where a 'diurnal averaged temperature' is needed.
The so-called diurnal averaged temperature can be obtained by solving the 1D
thermal conduction equation with a rotationally averaged energy conservation
condition as
\begin{equation}
(1-A_{\rm eff})\tilde{F}_{\rm s}=\varepsilon\sigma\tilde{T}^4-
\kappa\frac{\delta\tilde{T}}{\delta h}~,
\end{equation}
where $h$ means depth, $A_{\rm eff}$ is effective bond albedo derived from geometric
albedo, $\kappa$ is the thermal conductivity estimated from the above derived thermal
inertia, $\tilde{T}$ is the diurnal averaged temperature of interest, and
$\tilde{F}_{\rm s}$ refers to the diurnal averaged incident solar flux given by
\begin{equation}
\tilde{F}_{\rm s}=\frac{F_{\odot}}{r_{\odot}^2}\frac{1}{2\pi}\int^{2\pi}_{0}
\max(0,\vec{n}_{\odot}\cdot\vec{n}_{i}(\theta,\varphi)){\rm d}\varphi,
\end{equation}
in which $r_{\odot}$ is the heliocentric distance, $F_{\odot}$ is the solar constant
$1361.5\rm~Wm^{-2}$, $\vec{n}_{\odot}$ means the direction pointing to the Sun,
$\vec{n}_{i}(\theta,\varphi)$ is the normal vector of facet $i$ at the local
latitude $\theta$ and longitude $\varphi$.

\begin{figure*}
\includegraphics[scale=0.83]{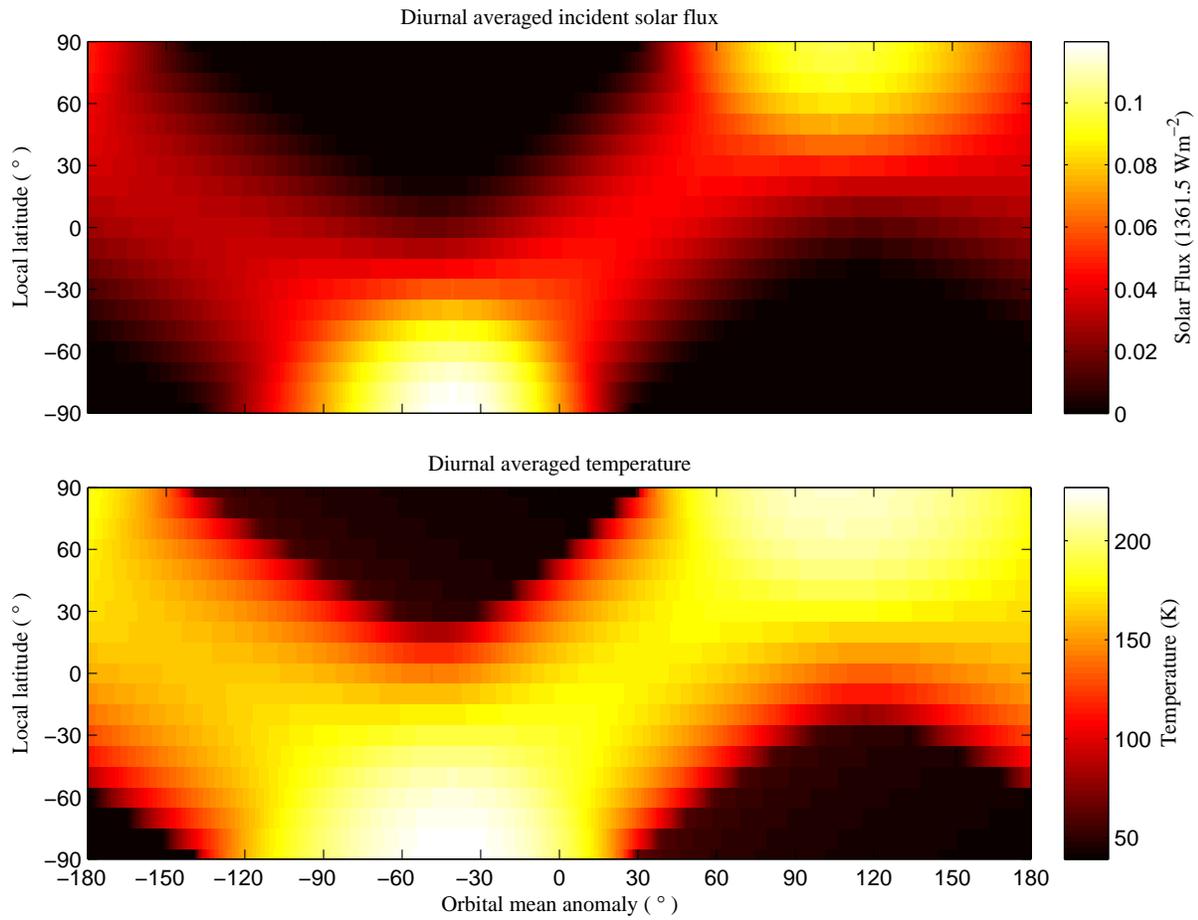}
  \centering
  \caption{Upper panel: the diurnal averaged incident solar flux $\tilde{F}_{\rm s}$;
  Under panel: the diurnal averaged temperature; horizontal axis stands for mean anomaly,
  where $0$ means the perihelion; vertical axis stands for local latitude on Dembowska,
  where $0$ means the equator and positive means north, negative means south. The surface
  temperature at each latitude shows significant seasonal variation, which may affect the
  surface properties of Dembowska.
  }\label{gst}
\end{figure*}

In Figure \ref{gst}, the upper panel shows the diurnal averaged incident solar flux
$\tilde{F}_{\rm s}$ on each latitude of Dembowska at different orbital position,
while the under panel shows the corresponding diurnal averaged temperature. We can
see that within an orbital period, the temperature changes smoothly at equator, but
shows large variations at high latitudes, where even appear polar night and polar day
at around mean anomaly$=-60^{\circ}$ and $=120^{\circ}$. Besides, at each orbital
position, the temperature of Dembowska's south and north region shows large difference,
which changes periodically following the orbital period. Thus it is reasonable to infer
that the thermophysical and spectral properties of the south and north region may be
different at different orbital position.

\begin{figure}
\includegraphics[scale=0.63]{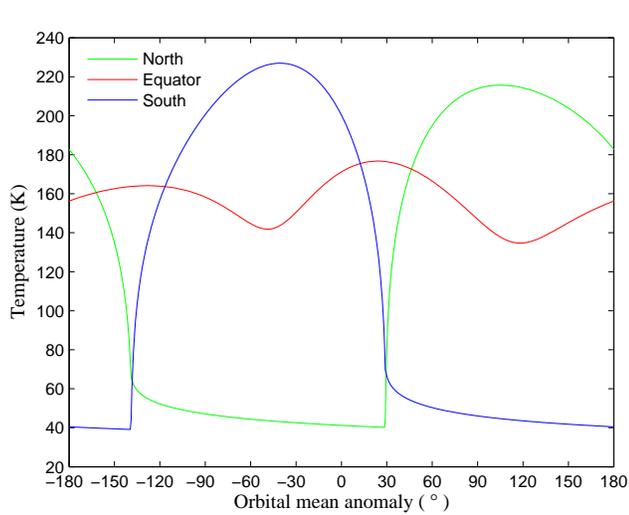}
  \centering
  \caption{Seasonal variation of the diurnal averaged temperature at equator, north and
  south region of Dembowska respectively.
  }\label{tempc}
\end{figure}

Figure \ref{tempc} further shows how temperature varies in an orbital period at equator,
north and south region of Dembowska respectively. We can see that the diurnal averaged
temperature at equator can fluctuate within $140\sim180$ K, while the temperatures at
north and south vary from minimum temperature $\sim40$ K to maximum temperature $\sim220$ K.
And, particular near anomaly$=-60^{\circ}$ and $=120^{\circ}$, the temperature difference
between the south pole and north pole can be as large as $\sim180$ K, which could cause
different thermophysical or spectral characteristics.

\begin{figure}
\includegraphics[scale=0.63]{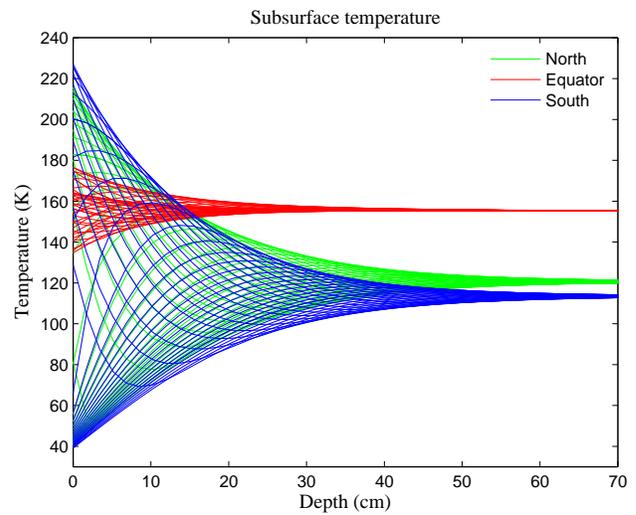}
  \centering
  \caption{Seasonal variation of the diurnal averaged temperature within the subsurface
  of Dembowska's equator, north and south region.
  }\label{subt}
\end{figure}

Figure \ref{subt} shows the temperature distribution within the subsurface
of Dembowska's equator, north, and south region. The curves labeled by red,
green, and blue stands for equator, north and south respectively. The different
curves plotted in the same color represents the temperature distribution at
different orbital position in a whole orbit period.
The subsurface temperature on the equator could achieve equilibrium $\sim160$ K
at about $30$ cm below the surface, while on the north and south region, the
equilibrium subsurface temperature appear to be $\sim120$ K at about $50$ cm depth.
Therefore, it is possible for us to further investigate the subsurface properties
by detecting the microwave emission of Dembowska at wavelength around $3\sim5$ cm.

\section{Discussion and Conclusion}
The radiometric method has been proved to be a powerful tool to determine
thermophysical properties of asteroids. In this work, we derive the thermophysical
characteristics of the large main-belt asteroid (349) Dembowska by using the
Advanced thermophysical model (ATPM) to reproduce the thermal infrared
data of Dembowska observed by IRAS, AKARI, WISE and Subaru, respectively.
The surface thermal inertia, roughness fraction, geometric albedo and effective
diameter of Dembowksa are well obtained in a possible 3$\sigma$ scale of
$\Gamma=20^{+12}_{-7}\rm~Jm^{-2}s^{-0.5}K^{-1}$, $f_{\rm r}=0.25^{+0.60}_{-0.25}$,
$p_{\rm v}=0.309^{+0.026}_{-0.038}$, and $D_{\rm eff}=155.8^{+7.5}_{-6.2}\rm~km$.

\begin{figure*}
\includegraphics[scale=0.83]{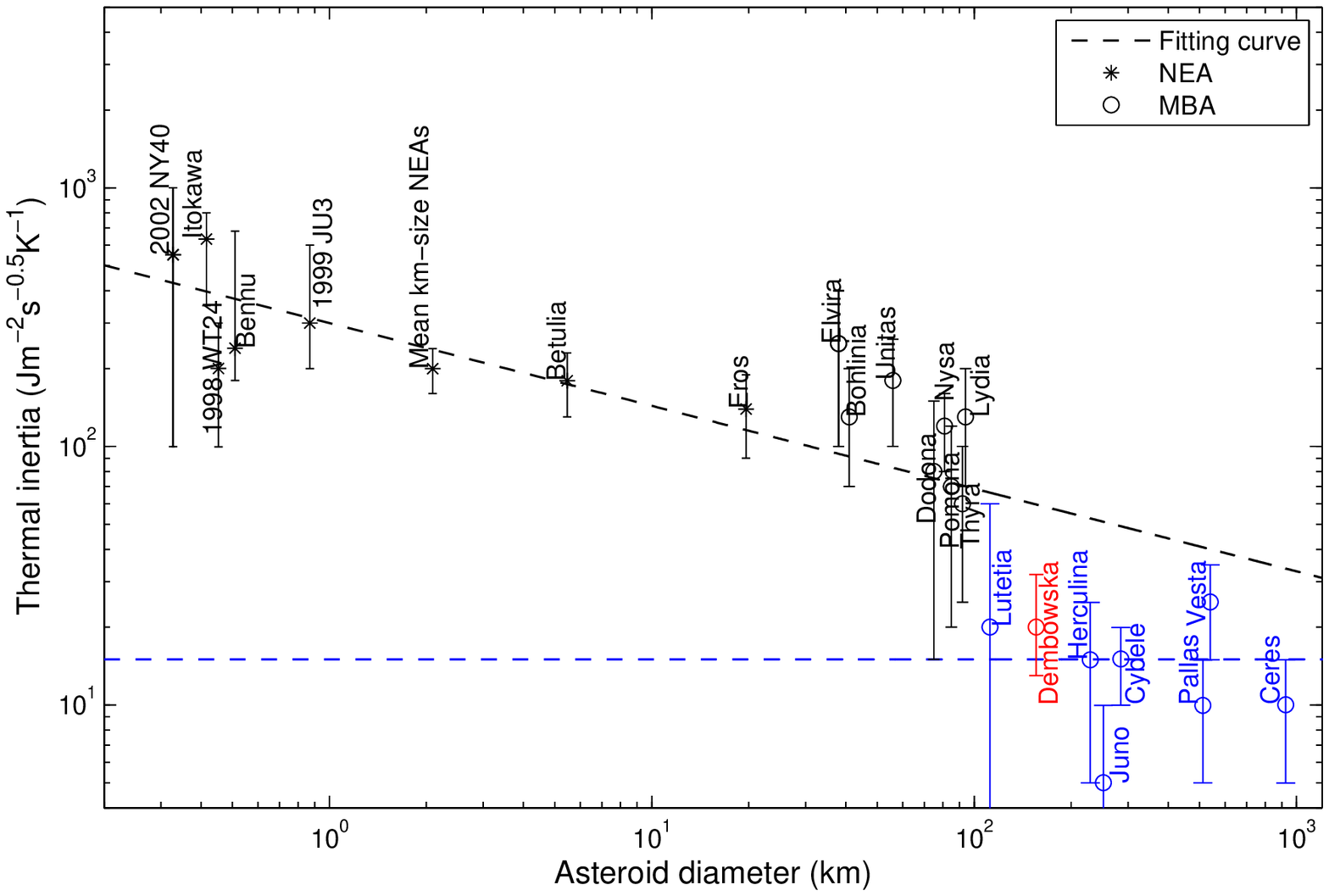}
\centering
\caption{Thermal inertia versus the size of asteroids \citep{Delbo2007,Delbo2009}.
  }\label{gammasize}
\end{figure*}

If we compare the thermal inertia and size of Dembowska derived in this work
with those of other asteroids that possess known thermal inertia and size given
in Figure \ref{gammasize} \citep{Delbo2007,Delbo2009},
we can find that the thermal inertia and effective diameter of the asteroids
$<100\rm~km$, e.g., NEAs or MBAs, will well follow the empirical
relationship given by \citet{Delbo2009}:
\begin{equation}
\Gamma=300D^{-0.32},
\end{equation}
which was originally used for NEAs.
However, for asteroids with diameters $>100\rm~km$, the thermal inertia does not show significant
dependence on their sizes, but seem to be all as low as about $15\rm~Jm^{-2}s^{-0.5}K^{-1}$.
Such interesting phenomenon indicates that larger asteroids (D $>100\rm~km$) might have
experienced long-lasting space weathering process and formed surface mantles without
disruption, which significantly reduced their surface thermal inertia. On the other hand,
the smaller asteroids (D $<100\rm~km$) might be the first or second generation impact
fragments and their surfaces have been repeatedly reshaped. \citet{Bottke2005} showed
that the asteroids with D $>100\rm~km$ are long-lived and only $\sim4$ out of $220$ disrupt
per Gyr, whereas most intermediate or smaller bodies (D $<$100 km) are fragments (or fragments
of fragments) created via a limited number of breakups of large asteroids with D $>$ 100 km.
Thus the dynamical lives of asteroids with D $>100\rm~km$ should be long enough to produce
a regolith layer with a sufficiently low thermal inertia.

The rotationally resolved data adopted in present work, uncovers a weak heterogeneous
feature between different local longitudes and latitudes of Dembowska. However due to
the absence of sufficiently precise observation data, the heterogeneous signal appears
to be rather weak to further infer how the surface characteristics differ on various
region. \citet{Abell2000} showed that Dembowska may be a heterogeneous body and suggested
that this asteroid may bear a large young impact crater. Nevertheless, such impact,
if happened, was on a relatively small-scale and would not alter the averaged thermal
properties of the global surface. Therefore, we may infer that the entire surface of
Dembowska should be covered by a dusty regolith layer with few rocks or boulders
on the surface.

On the other hand, we report that the surface temperatures on high latitudes of
Dembowska show large seasonal variations, as a result of the large axial tilt between the
rotational axis and orbital axis. This kind of seasonal variation can cause significant
temperature difference between the south and north region of Dembowska. Thus we argue
that the potential heterogeneous features between Dembowska's south and north region
might be induced by the seasonal effects or by large young impact crater. Further
investigations should be done to reveal this issue in future.

\section*{Acknowledgments}
The authors thank the reviewer Simon Green for the
constructive comments that improve the original
manuscript. We would like to thank Fumihiko Usui for providing the AKAKI data.
This work is financially supported by National Natural Science Foundation
of China (Grants No. 11473073, 11403105, 11661161013, 11633009), the Science
and Technology Development Fund of Macau (Grants No. 039/2013/A2, 017/2014/A1),
the innovative and interdisciplinary program by CAS (Grant No. KJZD-EW-Z001),
the Natural Science Foundation of Jiangsu Province (Grant No. BK20141509), and
the Foundation of Minor Planets of Purple Mountain Observatory.

\label{lastpage}

\end{document}